\documentclass[12pt]{article}
\def\frac#1#2{{\displaystyle#1\over\displaystyle#2}}

\usepackage[dvips]{graphics}
\usepackage[dvips]{epsfig}

\begin{document}
\author{A.~V.~Gurevich$^1$  and
K.~P.~Zybin$^1$.}
\title{High energy cosmic ray particles and the most
powerful new type discharges in thunderstorm atmosphere.}
\date{}
\maketitle

\begin{abstract}
The runaway breakdown -- extensive atmospheric shower discharge (RB - EAS)
excited in thunderstorm atmosphere by high energy cosmic ray particles
($\varepsilon_p>10^{17} - 10^{19}$ eV)  generate very powerful radio
pulse. The RB - EAS theory is compared with observations of radio pulses. An
agreement between the theory and experiment is established. The existence
of nowaday satellite and ground based systems which obtain regularly a large
amount  of observational radio data could allow to use them in combination
with other methods for effective study of high energy cosmic ray particles
\end{abstract}

\footnotetext[1]{P.~N.~Lebedev Institute of Physics, Russian Academy of
Sciences, 117924 Moscow, Russia.}

\section{Introduction}

Le Vine (1980) and Willet (1989) studying radio emission from thunderstorms
have noted a distinct class of powerful radio pulses. The intensive
investigations of this phenomena were performed during recent years.
For the measurements the satellite FORTE and specially constructed systems
as LASA, EDOT and others
were used [Smith et al., (1999),(2002); Thomas et al (2001);
Light and Jacobson (2002); Jacobson (2003)]. These studies allowed to establish
that the radio pulses having enormous power 100 -- 300 GW
are emitted in the wide frequency range by the intracloud
discharges in the upper troposphere.
The pulses are short time ($\leq 10\mu$s) and
have a definite bipolar form.  That is why they were called narrow bipolar
pulses (NBP). The observations show that these strong radio pulses are
accompanied by very weak optic emission only. A detailed analysis of the whole
complex of observational data allowed Jacobson (2003) to state that NBP is a
new type of thunderstorm discharge quite different from usual lightning. He
speculated that it could have relevance to runaway breakdown effect.

Runaway breakdown (RB) is a new physical concept of an avalanche type increase
of a number of relativistic and thermal electrons in air proposed by
Gurevich, Milikh and Roussel-Dupre (1992).  The avalanche can grow in electric
field $E\geq E_c$ which is almost an order of magnitude less than the
threshold of conventional breakdown. The electrons with high energies
$\varepsilon \geq (0.1 - 1)$MeV can become runaway and are
accelerated under the action of
electric field $E>E_c$. Directly this process -- acceleration and collisions
with air molecules lead to the avalanche type growth of the number of runaway
and thermal electrons (Gurevich and Zybin 2001).

Runaway breakdown in air (RB) is stimulated by the presence of a high
energy cosmic ray secondaries. Extensive atmospheric showers (EAS) are
accompanied by an effective local growth of the number
of cosmic ray secondaries
and thus have a strong influence on the RB process (Gurevich et al 1999).
The combined action of runaway breakdown and EAS  lead to the development of
RB - EAS discharge --  new type of electric discharge where relativistic
electrons  play a decisive role. RB-EAS discharge is accompanied by  strong
exponential growth of the number of energetic and thermal electrons,
positrons  and gamma quants (Gurevich et al 2004a). It can serve for the
generation of a strong bipolar radio pulse (Gurevich et al, 2002).

The goal of the present work is to extend a theory of RB - EAS discharge
in air to a very high energy range of
cosmic ray particles ($\varepsilon_p\geq 10^{17} -
10^{19}$ eV) and to compare the theory with the results of NBP observations.

\section{Narrow bipolar pulses (NBP)}

Narrow bipolar pulses are isolated short time discharges generated in
thunderclouds.
They were discovered and first studied by Le Vine (1980) and Willett
et al (1989). During last years very intensive and detailed studies of this
special type of atmospheric discharge were performed by Smith  et al (1999),
Rison et al (1999), Thomas et al (2001),
Light and Jacobson (2001), Smith et al (2003), Suszcynsky and Heavner (2003),
Jacobson (2003) and others. It was established that
NBP are  compact and energetic intracloud discharge.
They are observed in two forms: negative (NNBP)
and positive (PNBP). For negative first electric field peak is negative and
vice versa for positive.

{\bf Main features}

1. NBP are high altitude discharges. Their main location (Smith et al 2003)

z= 15 -- 20 km for NNBP (sharp peak near 18 km)

z= 7 -- 15 km for PNBP (sharp peak near 13 km)

2. Time characteristics of NBP (Smith et al 1999, 2002)

mean rise time  $\sim 1 - 2 \mu s$

full width at half maximum $\sim (2 - 5) \mu s$

full rise + fall time $\sim(5 - 10) \mu s$

3. NBP is observed as low frequency 0.2 -- 0.5 MHz bipolar ground
electromagnetic wave with large amplitude. The NBP effective field
amplitude  was measured simultaneously at several stations situated at
different distances  $R$ (Smith et al 2002, 2003).
Characteristic wave field is
$$
E\sim (10 - 30) \,
\left(\frac{100 km}{R}\right) V/m
$$

4. Electric current pulse, generating NBP is unipolar and
its maximum reaches values
$$
J_{m} \sim (30 - 100) kA
$$

5. Dipole electric moment change in the cloud determined by NBP discharge
is $M\approx 0.2 - 0.8$ Cu km (Smith et al 1999).

6. From the analysis of observational data it follows  that the NBP current
initiator moves with very high speed. The speed of initiator is determined
as $7.3\times 10^9 - 3.0\times 10^{10} cm/s$ (Smith et al 1999) or
$\sim 10^{10} cm/s$ (Jacobson 2003).

7. The NBP discharge rates is growing with the thunderstorm convection
strength (Suszcynsky and Heavner 2003).

{\bf HF radio emission}

1. NBP is always accompanied by intensive radio emission in a wide frequency
range up to 500 MHz  (Smith et al 1999).

2. Detailed study  of HF emission in the frequency range (26 -- 48) MHz
fulfilled at the FORTE satellite allowed to establish the following main
features (Jacobson 2003):

a. HF emission connected with NBP is always powerful its integrated ERP
$\geq 40$ kW It is called by Jacobson strong intracloud (IC) pulse.

b Strong IC pulses are {\it incoherent} in HF range.

c. Strong IC pulses are accompanied by very low
optic emission -- at least two orders of magnitude less than in usual flashes.

d. Strong IC pulses occur singly or  initiate intracloud flashes. They
never come within the interior of flashes. The optical emission does not
occur for initiator of strong IC pulse.

e. Strong IC pulse can occur without NBP, but not vice verse.

f. The strong IC pulse in a given storm appear to have a truncated ERP
distribution staying below a limiting ERP. For the most storms this is
of the order of 1 MW in the FORTE HF band 26 -- 48 MHz. The maximal
value of a limiting ERP is 10 MW.

\section{RB - EAS discharge stimulated by high energy CR particles.}

1. RB - EAS discharge is determined by the exponential growth of the number
of relativistic electrons, positrons and gamma quants. The characteristic
length $l_a$ is given in the Table for the main location heights of PNBP
(z=13 km) and NNBP (z=18 km). In RB - EAS discharge the RB process is
facilitated significantly in the energy range  3 -- 30 MeV due to effective
generation of gamma quants and $e^+ e^-$ pairs (Dwyer 2003,
Gurevich et al 2004a).

\begin{center}
Table: RB - EAS and NBP characteristics
\end{center}

%\begin{table}
\begin{center}
\begin{tabular}{|c|c c c|c c|}
\hline
$z(km) $ &  & RB-EAS theory  & & NBP observations & \\
\hline
&  $l_a$(m) & $\tau_g\,(\mu s)$ & $\tau_{rel}\,(\mu s)$ & $\tau_g\,(\mu s)$ &
$\tau_{rel}(\mu s)$  \\
\hline
13 & 200 & 0.7 & 0.7     &     1 - 2 & 2 - 5      \\
18 & 400 & 1.4 & 3     &     1 - 2 & 2 - 5      \\
\hline
\end{tabular}
\end{center}
%\end{table}

2. The production of a giant number of thermal electrons $n_{th}$
due to ionization of air molecules is going simultaneously with
the generation of fast relativistic electrons $n_{th} \sim 10^6 n_f$)
(Gurevich et al 2004). So the growth time of thermal and relativistic electrons
coincide. The attachment of thermal electrons to oxygen is due to three
particle collisions. The characteristic attachment time (Phelps 1969):
$$
\tau_{att}\sim\left(\frac{5\times 10^{18}}{N_m}\right)^2\,\mu s
$$
Here $N_m$ -- the number density of air molecules in cm $^{-3}$.
Attachment determines the relaxation time of discharge. The growth $\tau_g=l_a
/c$ and relaxation time of RB - EAS discharge are presented in the Table.

3. The electric current $J$ generated in RB - EAS discharge is unipolar.
For a given $E_m/E_c$ its maximum $J_m$ is proportional to the number of thermal
electrons. The last is proportional to the number of fast electrons and
thus proportional to the number of cosmic ray secondaries. That is why
the maximal current $J_m$  is proportional to the energy of cosmic ray particle
$\varepsilon_p$. For conditions
of strong RB - EAS discharge, using the calculations performed in
Gurevich et al 2004a, 2004b, we can estimate $J_m$ as
\begin{equation}\label{Jmax}
J_m\sim \left(\frac{\varepsilon_p}{10^{17}}\right)\,\hbox{kA}
\end{equation}
Here $\varepsilon_p$ is the energy of cosmic ray particle

4. The radio pulse emitted by RB - EAS discharge is bipolar. Its amplitude
depends on the distance to the source $R$ and on
discharge parameters. In average according to our simple model (\ref{Jmax})
it is proportional roughly to $\varepsilon_p$.

5. The main energy dissipated in RB - EAS discharge goes to
ionization of air molecules and to excitation
of $N_2$ vibration levels. Optic emission is very low -- less that 1\%
of total energy (Roussel-Dupre and Gurevich 1996, Gurevich et al 2004).

\section{Discussion}

{\bf NBP and RB - EAS theory}

Let us compare NBP with the theory of RB - EAS discharge generated at high
heights 13 -- 18 km by very energetic cosmic ray particles ($\varepsilon_p
\geq 10^{17} - 10^{19}$ eV). Parameters of RB - EAS discharge at high heights
are presented in the Table. One can see that both growth and relaxation
time characteristics roughly agree with NBP observations. Electric
current is unipolar. Its main peak reaches the values (\ref{Jmax}) which
generally are in agreement with the average data obtained in NBP observations.
One can see the following chain of measurements which show the rough
proportionality between $J_m$ and $\varepsilon_p$:

Tien Shang experiment (Gurevich et al 2004b)
$$
\varepsilon_p\sim 10^{15}\,\hbox{eV}\,,\qquad J_m\sim 1 - 10\, \hbox{A}
$$
Lightning first pulse measurements (Gurevich et al 2003)
$$
\varepsilon_p\sim 10^{16} - 10^{17}\,\hbox{eV}\,,\qquad J_m\sim 0.1 - 1\,
\hbox{kA}
$$
NBP
$$
\varepsilon_p\sim 10^{17} - 10^{19}\,\hbox{eV}\,,\qquad J_m\sim 10 - 100\,
\hbox{kA}
$$

Thus RB - EAS discharge can serve as background for explanation of NBP
phenomena.

Note that the high values of  maximal electric field
$E_m/E_c \approx 1.2 - 1.4$ were supposed in
estimates Gurevich et al (2004a). The direct observations of electric fields
at NBP heights gave lower values yet (Vonnegut et al 1989).
On the other hand the NBP are
usually seen in the active phase of storm near reflectivity core with 50
dBz radar reflectivity (Smith et al 1999).
According to Smith et al 2003 the strong positively charged
layer at the heights (15 - 16) km exist between generation
maximum of NNBP and PNBP. It is impotant also that  NBP
formation is accompanied by powerful HF emission.

{\bf HF emission model}

In Gurevich et al (2004a) RB - EAS model the air density $N_m$ distribution
was supposed to be smooth function depending on one coordinate $z$. In reality
in the high reflectivity core where NBP is generated a large number of
hydrometeors (liquid or frozen water drops) exist. That can generate
significant fluctuations in RB process. Really according to MacGorman and
Rust (1998) the characteristic hydrometeor dimensions in a high reflectivity
core $r_0\sim 0.3$ mm and density number $n\sim 10^{-2}$ cm $^{-3}$.
Due to this the mean free path of fast electrons between collisions with
hydrometeors is $l_{eh} \sim (\pi r_0^2 n)^{-1} \sim$ 0.3 km. The effective
radiation length of hydrometeors $t_h\approx 0.03 g/cm^2$ is equivalent to
the length $l_h = t_h/\rho_{air}\approx 1.5 - 3$ m in the air at the heights
13 - 18 km. It means that the passage of hydrometeor by runaway electron
is equivalent to the passage of additional air length $l_h$. Thus the existence
of randomly distributed hydrometeors lead to random fluctuations of $N_m$ and
$E_m/E_c$  of the order of $l_h/l_{eh}\sim$ 1\%.
Taking into account exponential dependence on $E_m/E_c$ of fast and thermal
electrons generated in RB process, one can expect that hydrometeors
could be the
reason of strong enough random fluctuations of electrons and electric current.

Another effect which could be responsible for fluctuations is space
inhomogeniety of cosmic ray secondaries which serve as a seeds of
RB - EAS process. These fluctuations are especially significant at the high
heights of the order 15 -- 20 km where the shower has not reach
the full stage of development yet. That is why the showers coming under
inclination angle more than $60^o$ are mostly effective
The approximation of a quasi flat electromagnetic shower cascade moving
together with cosmic ray particle which was used in Gurevich et al (2004a) could
be considered as a good approximation for low energy $\varepsilon_p<10^{15}$
eV only. For the higher energies of cosmic ray particles situation is quite
different. Note that laboratory experiments at accelerators
in this energy region are absent.
Both cosmic ray experiments and theory show very complicated character
of shower structure at high energies $\varepsilon_p\geq 10^{17}$eV
(Murzin 1988). The observations
using Pb and Pb-C cameras in Pamir and Chakaltaya experiments demonstrate
multidimensional and inhomogeneous structures of electron - photon shower
components (Baiburina et al 1984, Hasegawa and Tamoda 1996). Though photon
super families of halo type (Genina et al 1981) and a number of new type
processes "Centauro", "Chiron", penetrating chanels, anomalous transition
curves (see Gladysz -Dziaoshu 2001, Capdevielle and Slavatinsky 1999
) are not well understood, the
growth of fluctuations in photon and secondary electrons distribution
could exist. In spite of  fluctuations the part of the energy of
primary cosmic ray particle distributed in electron photon component
of the shower is conserved at the level $(0.2 - 0.3)\varepsilon_p$
(Baburina et al 1984). It
means that the full power of cosmic ray secondaries is proportional to
$\varepsilon_p$ what supports the basic relations (\ref{Jmax}).

Space and time fluctuations of $E_m/E_c$ and cosmic ray secondaries
lead to the corresponding fluctuations in number of generated thermal
electrons what results in a strong fluctuations of the current. In that a way
we suppose the intensive IC pulse HF radiation is generated.

At the same time these fluctuations do not change the essence of RB - EAS
process.  Its main features are reflected in IC HF radiation characteristics
emphasized by Jacobson (2003). Radio emission is strong but the accompanying
optic emission is very low, IC HF emission exists only as a single pulse
(or as a first pulse in the beginnings of a flash ) and so on.

{\bf Coherence effect and the giant power of low frequency radio emission
pulse (NBP)}

The power  dissipated in RB - EAS discharge is approximately proportional
to the energy of cosmic ray particle $\varepsilon_p$. The number of created
newborn  thermal electrons is also proportional to $\varepsilon_p$ and to
the exponential factor $F(E_m/E_c)$. Factor $F$ in real conditions could reach
values $F\sim 10^2 - 10^4$ (Gurevich et al 2003, 2004a).
Thus the full energy $W_d$ dissipated by
thunderstorm electric field to the creation of thermal electrons and hence to
the electric current of RB - EAS discharge is of the order of $\varepsilon_p
F$. For $\varepsilon_p = 10^{16}$ eV energy $W_d\sim 1$J, for $\varepsilon_p
\sim 10^{19}$ eV -- $W_d\sim 1$ kJ. The same is fully correct for a non coherent
HF radio emission. Its maximal frequency integrated power is 1--10 MW, maximal
emitted energy (10 -- 100) J.

The picture changes dramatically when we take into account {\it coherence of
low frequency (LF) emission process}. The pulse current region determined
by the EAS scale is $l_a\sim 300 - 400$ m. For LF radio emission (200 -- 500)
kHz the scale $L_c$ is of the order or less than the
length of radio wave $\lambda = (600 - 1500)$m. It means that the LF radio
emission is coherent process and due to this the {\it power of radio emission in
LF is growing with current $J$ proportionally to $J^2$}. That is why the ground
wave power in NBP could reach extreme values:
$$
P= \frac{2J^2}{3c} \,, \qquad\qquad P\approx (100 - 300)\,\hbox{GW}
$$
It means that emitted by radio pulse energy can reach 0.2 -- 1 MJ.
We see that the work of thunderstorm electric
field in NBP is really {\it giant}: it exceeds $10^6$ times the energy of
cosmic ray particle triggering the process.

We emphasize that the LF coherence effect is a strong argument supporting
RB - EAS model of NBP generation. LF coherence means that though HF emission
is incoherent but all currents which generate it
according to the model {\it has the same direction}.
In other words they have the same nature -- thermal electrons moving by
thundercloud electric field. These currents are only distributed in space -
time inhomogeneously in groups and thin filaments -- determined by
inhomogeniety of local electric field and  cosmic ray secondaries production.

{\bf Conventional ohmic electric discharge in the night time
ionosphere excited by NBP}

According to Smith et al (2003) in the night time conditions the NBP reflection
region lies at the hieghts $z$ 80 -- 90 km. One can see, that NBP power is so
high that electric field of the pulse $E_{nbp}$ coud reach and overcome the
value of conventional breakdown field $E_{th}(z)$ at these heights
$$
E_{nbp} > E_{th} = 22 \,\left(\frac{N_m(z)}{2.7\,10^{19} cm^{-3}}\right)\,
\frac{kV}{cm}
$$
where $N_m(z)$ is the number density of air molecules at the height $z$.
For the height 80 km it follows that $E_{th} = 24$ V/m, for the height 90 km
$E_{th} = 3.2$ V/m. Thus one can expect that short - time discharge (a few
$\mu$s) excited by NBP in lower ionosphere could be observed.

{\bf Statistics of high energy cosmic rays and NBP events}

The flux of cosmic ray particles having the energy
$\varepsilon>5\times 10^{18}$ eV is one  per km$^2$ per year (Murzin 1988).
The annual thunderstorm duration complied from 450 air weather system
in USA gives an average number 100 h/year (MacGorman et al 1984, Uman 1987).
It means that thunderstorms lasts about 1\% of the whole time.
The flux of high energy cosmic ray particles ($\varepsilon>5\times 10^{18}$eV)
in thunderstorm time is $0.01/km^2year$. The FORTE satellite combined with
ground based systems allows to detect simultaneously radio emission
at $(1\div 3) 10^6$  km$^2$.  Thus the number of events
of interaction of cosmic rays  $\varepsilon>5\times 10^{18}$eV
with thunderclouds is $(1\div 3) 10^4$ per year. In reality if
the NBP can be generated by lower energies say $5\times 10^{17}$ the statistics
would be 100 times better $\sim 10^6$ events per year. Thus we see that there
is no contradictions between the NBP observational data ($10^5$ events during
4 years (Smith et al 2003)) and high energy cosmic ray flux.

{\bf Initiator speed}

According to observations (Smith et all 1999, Jacobson 2003) the speed of NBP
initiator is close to $10^{10}$ cm/s. The high energy cosmic ray particle
(HECRP) moves with velocity of light. But the radio emission of RB -EAS
discharge is generated by the current of the thermal electrons which flows in
the direction of  thunderstorm electric field {\bf E}. Thus the speed of
initiator $v = c\,cos\,\theta$, where $\theta$ is the angle between {\bf E} and
HECRP direction of motion. That explains the difference in initiator speeds.
We emphasize that values of observed velocity is not far from $c$.

We note that for effective development of shower the cosmic ray particle has to
pass a thick enough atmospheric layer. At the heights 13 -- 18 km the density
of air is low. That is why HECRP has to go under the large inclination angles
$\theta \approx 70^o - 80^o$ to obtain well developed shower. Thus the RB - EAS
discharge become stronger for HECRP having large inclination angle.

{\bf High energy cosmic ray particle detection (preliminary remarks)}

The high power of NBP signals allows to detect them at large distances --
up to 1000 km. That can be used for detection of cosmic ray particles with
energies $\varepsilon_p\sim 10^{18} - 10^{19}$ eV and higher (HECRP). Of course
every concrete event depends not only on $\varepsilon_p$ but on a number of
others factors like $E_m/E_c$, reflectivity core size, thunderstorm activity.
But one can expect that after averaging over a large number of events some
useful information about HECRP could be obtained directly from NBP. For
example, according to Smith et al (2003) the NBP database  has now more 100 000
events.  NBP is created by current having maximal value $J_m$ which is
proportional to $\varepsilon_p$ (\ref{Jmax}). As the HECRP integral flux
$F(\varepsilon>\varepsilon_p)$ is proportional
to $\varepsilon_p^{-2}$ one can construct the averaged over all NBP events
function
$$
Q=\left< \Phi(J_m>I_0)\left(\frac{J_m}{I_0}\right)^2\right>
$$
Here $\Phi$ is the full number of NBP with $J_m> I_0$ and
$I_0$ is arbitrary chosen current, say  5 kA.  A careful selection of PNBP
and NNBP homogeneous conditions
using HF emission data should be performed. At the middle values $J_m< 100$ kA
the function $Q(J_m)$ is expected to be approximately constant
what  will show that cosmic ray spectrum plays a dominant role. In that case
the curve $Q(J_m)$ and its behavior at the highest values $J_m\geq 300$ kA can
give evidence of the {\it existence of GZK cutoff or its absence}.

Here we supposed that up to highest
values of $J_m$  flux dependence on $\varepsilon_p$
is decisive.  The role of other thunderstorm factors should be carefully
studied.

\section{Conclusion}

The analysis presented in the letter allows to formulate the following
main statements:

1. Narrow bipolar radio pulses (NBP) are generated by runaway breakdown
-- extensive atmospheric shower (RB -EAS) discharge initiated in atmosphere by
very high energy cosmic ray particles.

2. The detailed analysis of a large NBP data base could be used to obtain
unic information about HECRP spectra including a highest energy range. There
is a chance to establish the fundamental fact of existence  (or non existence)
of GZK cutoff.

3. Of special interest is the detailed analysis of giant information gathered
in HF radio spectrum observations. These new data analysis being combined with
the theory and ground based measurements give a chance to reach some progress
in understanding of basic processes in high energy particle physics
($\varepsilon > 10^{16}$ eV).

{\bf Acknowledgements}

The authors are grateful to Prof. G.T.Zatsepin, Prof V.L.Ginzburg, Prof.
E.L.Feinberg, Prof. O.G.Ryajzkaya, Dr. H.C.Carlson, Dr. L.M.Duncan, Dr. V.S.
Puchkov, Dr. A.P.Chubenko for discussion.

The work was supported by EOARD-
ISTC grant \#2236, ISTC grant \#1480, by the President of Russian Federation
Grant for Leading Scientific Schools Support
and by the Russian Academy Fundamental Research Program "Atmosphere Physics:
Electric Processes, Radio Physics Methods".

\newpage

{\bf References}

Baburina S.G.,Borisov A.S., Guseva Z.M., et al  Transaction
of P.N.Lebedev

Physical Institute (FIAN) {\bf 154}, 1 - 142, (1984); Nucl. Phys. B {\bf 191},
1, (1981)

Capdevielle J.N., S.A.Slavatinsky  Nucl. Phys. B {\bf 751}, 12, (1999)

Dwyer J.R. Geophys. Res. Lett. {\bf 30}, 2055 (2003)

Genina L.E., J.E. Vasinyuk, A.S. Nanasian et al., Preprint FIAN N54, (1981)

Gladysz - Dziades  E., Preprint Institute of Nuclear Physics, Crakow (2001)

Gurevich A.V., G.M. Milikh, R.A.Roussel - Dupre Phys.Lett. A 165, 463 (1992)

Gurevich A.V., K.P.Zybin  Physics Uspekhi {\bf 44}, 1119, (2001)

Gurevich A.V., K.P.Zybin, R.A.Roussel - Dupre Phys.Lett. A 254, 79 (1999)

Gurevich A.V., L.M. Duncan, Yu.V.Medvedev and  K.P.Zybin,

Phys.Lett. A 301,
320 (2002)

Gurevich A.V., L.M. Duncan, A.N.Karashtin and  K.P.Zybin,

Phys.Lett. A 312,
228 (2003)

Gurevich A.V., Yu.V.Medvedev and  K.P.Zybin,  Phys.Lett. A 321, 179 (2004)

Gurevich A.V., Yu.V.Medvedev and  K.P.Zybin, Phys.Lett. A
in preparation (2004a)

Gurevich A.V., A.N.Karashtin, A.P.Chubenko, et al
hep-ex / 0401037,

Phys.Lett. A  (2004 b)

Hasegava S., M.Tamada Nucl. Phys. B {\bf 474}, 141 (1996)

Jakobson A.R., J. Geophys Res. {\bf 108}  4778 (2003)

LeVine D., Journ. Geophys. Res., {\bf 85}, 4091 (1980)

Light T.E., A.R. Jakobson J.Geophys. Res. {\bf 85} 4756  (2002)

Mac Gorman D.R., M.W. Maier, W.D.Rust NUREG/ CR - 375, Nuclear Regulatory

Comission, Washington (1984)

Mac Gorman D.M., W.D.Rust "The electric nature of storms" N.Y. 1998

Murzin V.S., "Physics of Cosmic Rays"  Moscow University Press (1988)

Phelps A.V., Can. J. Chem. {\bf 47}, 1783, (1969)

Rison W., R.J.Thomas, P.R.Krehbiel, T.Hamlin, J.Harlin,

Geophys. Res. Lett. {\bf 26}, 3573, (1999)

Roussel - Dupre R.A., A.V.Gurevich J.Geophys.Res {\bf 101}, 2297, (1996)

Smith D.A., X.H.Shao, D.N.Holden et al. J.Geophys.Res. {\bf 104}, 4189 (1999)

Smith D.A., M.J.Heavner, A.R.Jacobsen, et al, Preprint LANL (2002)

Smith D.A., K.B.Eack, J.Harlin et al, J.Geophys.Res. {\bf 107}, 4183, (2002)

Suszynsky D.M., M.J.Heavner Preprint LANL (2003)

Uman M.B., The lightning Discharge, Academic Press (1987)

Vonnegut B., Blakeslee R.J., H.C.Christian J.Geophys.Res. {\bf 84}, 13135,
(1989)

Willet J.G., J.C.Bailey, E.P.Krider J.Geophys.Res. {\bf 94}, 16255 (1989)

\end{document}